\documentclass[aps,prl,reprint,superscriptaddress,twocolumn]{revtex4}

\usepackage{textcomp}
\usepackage{graphicx}

\begin{document}

\title{Fragile 3D Order in V$_{1-x}$Mo$_x$O$_2$}

\author{Matthew A. Davenport}
\affiliation{Department of Chemistry and Biochemistry, The University of Alabama, Tuscaloosa, AL 35487, USA}
\author{Matthew J. Krogstad}
\affiliation{Materials Science Division, Argonne National Laboratory, Lemont, IL 60439, USA}
\author{Logan M. Whitt}
\affiliation{Department of Chemistry and Biochemistry, The University of Alabama, Tuscaloosa, AL 35487, USA}
\author{Chaowei Hu}
\affiliation{Department of Physics and Astronomy and California NanoSystems Institute, University of California, Los Angeles, Los Angeles CA 90095, USA}
\author{Tyra C. Douglas}
\affiliation{Department of Chemistry and Biochemistry, The University of Alabama, Tuscaloosa, AL 35487, USA}
\author{Ni Ni}
\affiliation{Department of Physics and Astronomy and California NanoSystems Institute, University of California, Los Angeles, Los Angeles CA 90095, USA}
\author{Stephan Rosenkranz}
\affiliation{Materials Science Division, Argonne National Laboratory, Lemont, IL 60439, USA}
\author{Raymond Osborn}
\affiliation{Materials Science Division, Argonne National Laboratory, Lemont, IL 60439, USA}
\author{Jared M. Allred}
\email[]{jmallred@ua.edu}
\affiliation{Department of Chemistry and Biochemistry, The University of Alabama, Tuscaloosa, AL 35487, USA}

\date{\today}

\begin{abstract}
The metal-to-insulator transition (MIT) in rutile VO$_2$ has proven uniquely difficult to characterize because of the complex interplay between electron correlations and atomic structure. Here we report the discovery of the sudden collapse of three-dimensional order in the low-temperature phase of V$_{1-x}$Mo$_x$O$_2$ at $x=0.17$ and the emergence of a novel frustrated two-dimensional order at $x=0.19$, with only a slight change in electronic properties. Single crystal diffuse x-ray scattering reveals that this transition from the 3D M1 phase to a 2D variant of the M2 phase results in long-range structural correlations along symmetry-equivalent (11L) planes of the tetragonal rutile structure, yet extremely short-range correlations transverse to these planes. These findings suggest that this two-dimensionality results from a novel form of geometric frustration that is essentially structural in origin.
\end{abstract}
\maketitle
In VO$_2$, the V$^{4+}$ $d^1$ electrons are metallic at high temperature (HT), but localize into a paramagnetic semiconducting state at low-temperature (LT), below a first-order structural phase transition at 340 K. Electron-electron correlations are considered by many to be responsible for this transition, possibly in combination with a Peierls instability \cite{Haverkort2005,Weber2012MottPeierls,Holman2009,Rice1992,BritoDMFT2016,Huffman2017}. Others still favor the primacy of a purely structural instability, suggesting previous calculations underestimated the effect of bond covalency, entropy, and/or orbital ordering \cite{Budai2014,Wall2018,Hiroi2015,Xu2017}. Sixty years after the discovery of this metal-insulator transition \cite{Morin1959}, there is no consensus concerning its origin.
 
\begin{figure}
	\includegraphics[width=0.5\textwidth]{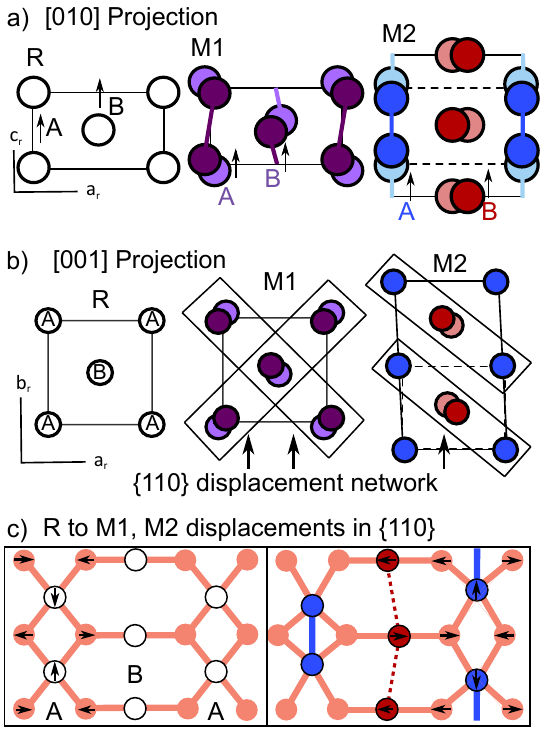}
	\caption{\textbf{a} $[010]_{\rm{R}}$ projection of the R, M1, and M2 structures. The same three structures are shown in \textbf{b} from the $[001]_{\rm{R}}$ projection. Boxes are drawn around $\{110\}_{\rm{R}}$ distortion planes. Oxygen atoms are omitted from \textbf{a} and \textbf{b} for clarity. \textbf{c} Step-wise schematic showing the conventional understanding of dimer formation driving the long-range distortion in two dimensions within $\{110\}_{\rm{R}}$. The distortion occurs along one set of planes in M2 and two sets of superposed planes in M1.\label{structure}}
\end{figure}

Whatever the role of electronic correlations, the structural instability itself is more complex than first perceived. The rutile structure of metallic VO$_2$ (R) has two chains (A and B) of edge-sharing octahedra along the $c$ axis, which is shown viewed across and down the chains in Fig. \ref{structure}a and b, respectively. Also shown are the LT M1 and M2 phases. In the M1 phase of pure VO$_2$, every metal atom dimerizes below the MIT, causing the neighboring chains to buckle in orthogonal directions: the A chain, along [1,1,0] and the B chain, along [1,$\overline{1}$,0]. The M2 phase is similar to the M1 phase, except that only half the metal atoms dimerize. That is, if the A chain dimerizes, only the B chain buckles \cite{Hiroi2015,DHaenens1975,villeneuve:1973}. Simple Coulombic arguments can explain the connection between dimerization and buckling, illustrated in Fig. \ref{structure}c. The $c$-axis displacements inherent to $M$-$M$ bond formation in one chain couple to the $ab$ displacement of the metal at [$\frac{1}{2}$,$\frac{1}{2}$,$\frac{1}{2}$] in the next chain through the shared oxygen atom (Fig. \ref{structure}c). A 2D network propagates in the (110) planes, with two types of V atom displacements in the A and B chains, respectively. The M1 phase can be seen as a superposition of two orthogonal M2 networks. The fact that the M2 phase is semiconducting, even though only half the V ions dimerize, has been interpreted as evidence that the MIT is not a pure Peierls instability \cite{Rice1992}. The M2 phase has been observed in pure VO$_2$ when strained or in a thin film \cite{Kim2016,Okimura2012}, but not in the bulk material \cite{CorrPDF2010} unless substituted with an ion such as Cr$^{3+}$ \cite{villeneuve:1973}. Additionally, the electronic transition in VO$_2$ thin films can be decoupled from the structural one according to a number of recent reports \cite{Yang.2016,Lee:2018ch,Ji.2021,Laverock:2014}. These thin films and heavily strained crystals were described as having a ``monoclinic- like metallic'' state at low temperature, since a detailed structural characterization was not possible and the symmetry can only be inferred.

Several attempts have been made to more fully describe the connection between the M1 and M2 phases in tandem \cite{BritoDMFT2016,Lovorn2017}. A phenomenological approach was taken by Lovorn and Sarker, who used the Ashkin-Teller model using displacements \textit{in lieu} of spins \cite{Lovorn2017}. They found that the 2D displacement network shown in Fig. \ref{structure}c can only become fully three-dimensional by including second-order terms to the Hamiltonian arising from weak cooperative displacements in the $\langle 100\rangle$ directions. In other words, the structure with only nearest-neighbor interactions is geometrically frustrated, and 3D order is only stabilized by next-nearest neighbor interactions. This model would be generalizable to any dimerized rutile system, making it relevant to any detailed structural study.

This work aims to better understand the structural instability in VO$_2$ by perturbing the bulk transition. We have found a new manifestation of short-range ordering that emerges when the M1 phase is suppressed, providing strong experimental support for the Lovorn-Sarker model. The substitution of $d^2$ Mo$^{4+}$ ions into VO$_2$ is known to suppress the MIT monotonically until $\approx 30$\% Mo, while also enhancing the metallicity of the LT phase \cite{Holman2009}. Here we show that between $x=0.17$ and $x=0.19$ the LT structure associated with the electronic transition instead transforms from a fully 3D-ordered M1 phase into a short-range-ordered M2 phase, which has extended two-dimensional correlations within (110) planes and much weaker transverse correlations, together indicating a high degree of geometric frustration. Our transport measurements show that the electronic transitions into the new ``2D-M2'' phase at $x = 0.19$ and the regular M1 phase at $x = 0.17$ are virtually indistinguishable, in spite of the differences in the structural transitions, which suggests a connection to previous measurements on thin films and strained crystals in the parent VO$_2$ phase.

This 2D-M2 ordering was revealed using three-dimensional difference pair distribution functions (3D-$\Delta$PDF). 3D-$\Delta$PDF is a recently developed method that builds substantially on the powder-PDF method \cite{Billinge2008,welberry2016}. Large 3D volumes of single crystal x-ray scattering data are transformed into reciprocal space coordinates, after which the Bragg peaks can be removed using the ``punch-and-fill'' method \cite{Weber2012pdf}, leaving only diffuse scattering. A Fourier-transformation of this diffuse scattering gives a PDF map that only contains information about the difference (hence the $\Delta$) between short-range and long-range ordering. Positive (negative) peaks in real-space occur at interatomic vectors that correspond to increased (decreased) electron density compared to the average long-range structure. From this, features such as atomic displacements are visually apparent.

\begin{figure}
	\includegraphics{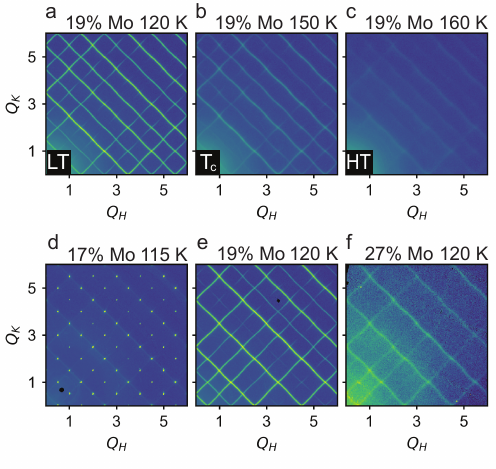}
	\caption{Reciprocal lattice slices from V$_{1-x}$Mo$_x$O$_2$. \textbf{a-c} The $L$ = 0.5 slices of 19\% Mo, at $T$ = 120, 150, and 160\,K, respectively. \textbf{d-f} Composition dependence of the $L$ = 1.5 LT structures at 17,19, and 27\%. All coordinates are given in reciprocal lattice vectors of the parent R phase.\label{rspace}}
\end{figure}

Total x-ray scattering was measured on individual, single crystals from V$_{1-x}$Mo$_x$O$_2$\cite{supplemental} rotated over 360\textdegree, and then used to reconstruct 3D data in reciprocal space coordinates. Cuts from such reconstructions are shown on a log scale in Fig. \ref{rspace}. Intense, sharp rods of scattering were observed below 150\,K in $x$ = 0.19 (19 \% Mo) at $L$ = half-integer planes. The high-resolution of the measurement resolves subtle features such as a mild off-axis curvature in the scattering rods. The rods also show a consistent profile and intensity across the full \textbf{Q} range measured, indicating that the correlations are very strong even down to a fraction of an atomic radius. Fig. \ref{rspace}a-c show the $T$-dependence of the ($HK\frac{1}{2}$)$^*$ plane in 19\% Mo through the structural transition at $T$ = 150\,K. Compositional dependence is shown in Fig. \ref{rspace}d-e for Mo compositions of 17,19, and 27 \%, respectively. The 17\% Mo crystal exhibits the normal structural phase transition to the M1 phase at 168\,K, although very weak rods can be observed between the super-cell Bragg peaks. The 27 \% Mo sample shows the same network of rods as the 19\% one, except that the higher composition shows a drastically reduced intensity and increased peak width.

\begin{figure}
	\includegraphics[width=0.45\textwidth]{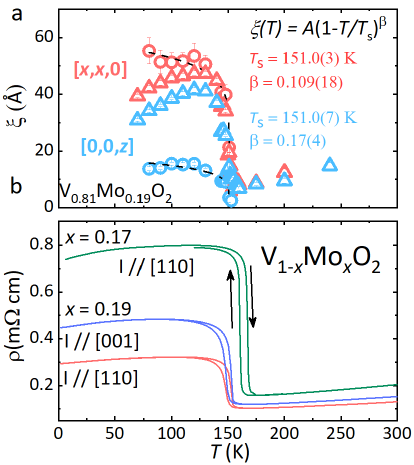}
	\caption{\textbf{a} $\xi (T)$ in 19\% Mo as determined from fitting the scattering rod width (circles) and from the 3D-$\Delta$PDF (triangles). Red and blue symbols are for the [$x,x,0$] and [$0,0,z$] directions, respectively. \textbf{b} $\rho(T)$ of selected orientations from 17\% and 19\% Mo crystals.\label{params}}
\end{figure}

The cross-section of the scattering rods in the $[HH0]^*$ and $[00L]^*$ axes can be fit to a Pseudo-Voigt function. The correlation length, $\xi_i$ was calculated from the Lorentzian component using the expression $\xi = \frac{1}{\Gamma_i} $ where $\Gamma_i$ is the Lorentzian full width at half maximum in reciprocal space coordinates (\r{A}$^{-1}$) $i$ is the axis of the cut. Fitting the $T$ dependence of $\xi$ to a power law $\xi=A(1-\frac{T}{T_s})^\beta $ gives a transition temperature of 151.0(3)\,K and a critical exponent of $\beta$ = 0.109(18) according to the $[HH0]^*$ cross-section. The resulting $\xi_{xx}$ and $\xi_z$ are plotted in Fig. \ref{params}a as red and blue circles, respectively. The power law behavior suggests that this first-order phase transition is quasi-2D, since hysteresis and volume discontinuities rule out a second-order phase transition. The maximum intensity of the scattering rods occurs around 120\,K. The structural properties are compared to the electronic ones using electronic transport, which is shown for the $x = 0.17$ and 0.19 in Fig. \ref{params}b. These two compounds are metallic in both HT and LT phases with the resistivity, $\rho$ in the m$\Omega$-cm range. The sudden increase in $\rho(T)$ by a factor of 3 to 8 at the structural transition suggests the formation of charge gaps, and is reminiscent of a charge density wave metal.

\begin{figure}
	\includegraphics{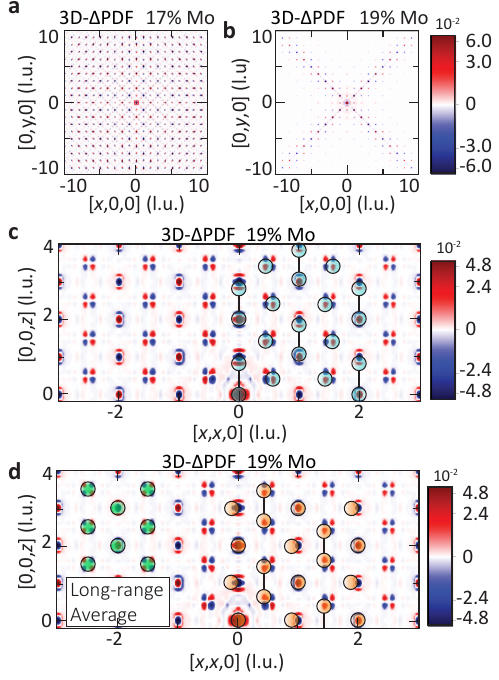}
	\caption{Symmetric log plots of the 3D-$\Delta$PDF correlation maps from \textbf{a} 17\% Mo and \textbf{b} 19\% Mo at 115 and 120\,K, respectively, parallel to the (001) plane at $z$ = 0.  A slice from the 19\% Mo (110) plane are shown in \textbf{c} and \textbf{d}. Blue and orange circles show the intratomic vectors of metal sites relative to a dimerized or buckled site at the origin, respectively.The green circles show the same in the undistorted rutile structure. \label{fft}}
\end{figure}

Cuts from the 3D-$\Delta$PDF are shown in Fig. \ref{fft}a-d. The 17\% Mo phase again serves as a standard for usual 3D M1 ordering (Fig. \ref{fft}a), with long-range correlations along all crystallographic axes. This is in contrast to the unusual two-dimensionality observed in 19\% Mo (Fig. \ref{fft}b). In the latter, the strongest correlations fall along (110) and (1$\overline{1}$0) planes, as expected from the scattering rod orientations. The two crossed planes each come from different nanoscale domains that chose different ordering axes.

Despite the complex pattern in the PDF, the simple structure of undistorted rutile allows an unambiguous assignment of most of the observed features (Fig. \ref{fft}c,d). The pattern of positive and negative peaks at integer points, [$u,u,w$], indicates the local unit cell is doubled in this plane. The quadrupolar features centered at [$\frac{u}{2},\frac{u}{2},\frac{w}{2}$] represent interchain metal-metal correlations. Weaker dipolar features are observed around the points [$\frac{u}{3},\frac{u}{3},w$], [$-\frac{u}{3},-\frac{u}{3},w$], [$u+\frac{1}{6},u+\frac{1}{6},\frac{w}{2}$], and [$u-\frac{1}{6},u-\frac{1}{6},\frac{w}{2}$], corresponding to correlations involving oxygen atoms. Also, the same pattern is repeated in weaker planes of correlations beyond the principal \{110\} plane.

The (110) cut of the 3D-$\Delta$PDF map is equivalent to a 2D cut of the M2 phase on the (110)$_R$ (Fig. \ref{structure}c) and is inconsistent with the M1 structure (See Supp. Mat. for more details). The correlations are very strong in the actual compound, and remain measurable in the 3D-$\Delta$PDF at least to 25 unit cell diagonals (about 20\,nm), but die-off within just a few unit cells ($\approx$1\,nm) in the orthogonal axis.This structural model will be referred to as the `2D-M2' model to emphasize the 2D nature of the coherent displacements. The 3D-$\Delta$PDF map was used as an alternate method for estimating $\xi$, by fitting the PDF amplitudes to a back-to-back exponential decay function \cite{Krogstad:2019tc}. The results are plotted in Fig. \ref{params}a (triangular symbols) along with the fits from scattering rod widths. Notably, the $[x,x,0]$ length scale agrees very closely for both methods, while the $[0,0,z]$ seems to be underestimated by the pseudo-Voigt fits.

\begin{figure}
	\includegraphics[width=0.48\textwidth]{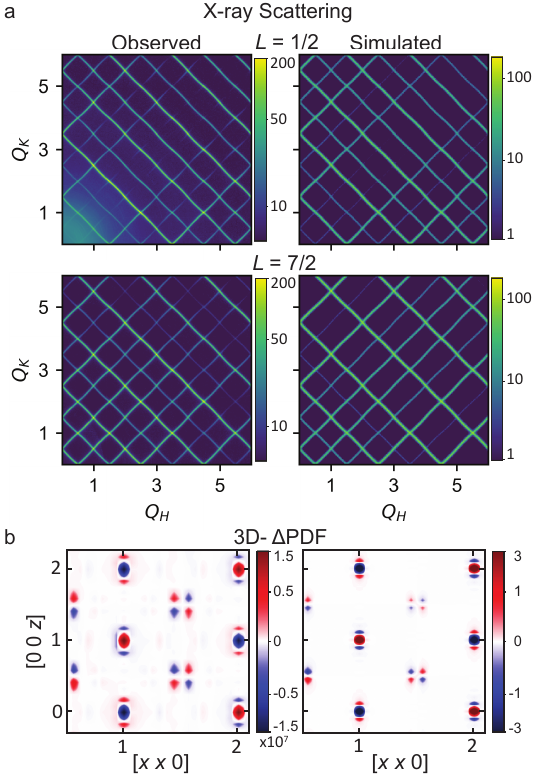}
	\caption{\textbf{a} Comparison of observed diffuse scattering from various slices, on left, to the simulated scattering model on right. \textbf{b} Comparison of observed 3D-$\Delta$PDF, on left, to the calculated disorder model, on right.\label{models}}
\end{figure}

In order to test the 2D-M2 interpretation, metal-only 2D-M2 models were constructed for both the x-ray scattering and 3D-$\Delta$PDF \cite{Paddison2019,Simonov2014,DISCUS}. Fig. \ref{models} compares the results of these simulations to observations, which match closely. It is clear that the constructed 2D-M2 model naturally conforms to observation, even down to the relative intensities of the scattering rods. Note that one extra parameter, ferrodistortive correlations along the [100] and [010] axes, was required in the diffuse scattering simulations to reproduce the wave-like bending of the scattering rods. The weak ferrodistortive correlations are frustrated by the main in-plane correlation, which is discussed further below. The calculated 3D-$\Delta$PDF map reproduces both the integer and half-integer lattice vector features. The weaker features attributed to oxygen atoms are missing, as expected. 

The constructed models validate the 2D-M2 interpretation of the data. Combined with the other measurements we made, this shows that the 2D-M2 is a different thermodynamic phase from both the long-range M1 and R phases. The transition from 2D-M2 to R is a first-order, metal-to-metal phase transition. Compositionally, the structural instability in V$_{1-x}$Mo$_x$O$_2$ changes from M1 to 2D-M2 ordering quite discontinuously between $0.17 < x \leq 0.188$. The change is very similar to the observed V$_{1-x}$Cr$_x$O$_2$ phase behavior, except that the Cr M2 analog has always been reported to be ordered in three dimensions (3D-M2). Another difference is that Cr-substitution immediately ($\approx$1.5\%) suppresses M1 in favor of 3D-M2.

All of this is reminiscent of the ``embedded 2D crystal model'' proposed by Lovorn and Sarker \cite{Lovorn2017}. In accordance with the Lovorn-Sarker model, the critical composition near 19\% Mo is when the orthogonal, inter-penetrating order parameters become inequivalent ($m_A\neq 0\neq m_B$). Geometric frustration between the $\langle 110\rangle$ and $\langle 100 \rangle$ directions are then strong enough to prevent ordering along both directions simultaneously. Instead, regions with strongly coupled displacements in either (110) or (1$\overline{1}$0) planes form, as observed here. Unexpectedly, the inclusion of the frustrated $\langle 100\rangle$ correlations was required to explain the wavy component in the scattering rods in the present diffraction data, making the frustration a necessary part of the model. To our knowledge, geometric frustration of displacements inhibiting ordering in a single dimension has never been observed experimentally in rutile or any other extended solid. 

We now consider the broader implications of these observations on the structural instability in rutile. VO$_2$ and MoO$_2$ show structurally identical HT and LT phases, though MoO$_2$ has a double bond between metals instead of a single one, and $T_s$ increases from 340\,K to 1400\,K, both likely owing to the extra $d$-electron. In this light, the substitution of Mo into VO$_2$ ought to enhance $T_s$, in contrast to the observed rapid suppression. The addition of geometric frustration to this picture resolves this inconsistency, wherein relative strengthening of certain interactions enhances frustration and suppresses 3D ordering \cite{Lovorn2017}. That is, VO$_2$ itself is already strongly frustrated, the $T_{MIT}$ is already suppressed, and the M1 structure is on the brink of collapse \cite{Hiroi2015}. From an electronic perspective, metallicity is retained in the LT structures of V$_{1-x}$Mo$_x$O$_2$ near the critical composition $x = 0.19$ for both the M1 phase in $x$ = 0.17 and the 2D-M2 phase in $x$ = 0.19 (Fig. \ref{params}b), which suggests that the electronic ordering is not fundamentally altered between M1 and 2D-M2. It is only the long-range structural manifestation of those electronic effects that changes.

These considerations are reminiscent of strained crystals or thin films of VO$_2$, where electronic ordering is decoupled from the structural changes. Mo's larger \textit{4d}-orbitals may initially enhance frustration via additional overlap along the $\left\langle 100\right\rangle$ axes. We suggest that geometric frustration is similarly in pure VO$_2$ when strained or in thin films, where 2D confinement and loss of cohesive forces affect bond geometries. \cite{Kim2016,Ji.2021,Lee:2018ch,Yang.2016,Arcangeletti:2007}. In fact, among the various reported `monoclinic-like metallic' phases, one was recently observed in VO$_2$ thin films that featured stripe formation oriented along the [110]$_R$ axis \cite{Laverock:2014}, which are all features consistent with the 2D-M2 phase characterized here. A direct relationship has not yet been established, but if the 2D-M2 and monoclinic-like metallic states are indeed the same phases, then the present study would be the first detailed structural characterization, owing to its formation in a bulk single crystal under ambient pressure. The precise structural details reported enabled the first experimental verification of geometric frustration in VO$_2$ transition.

A new type of ground state in V$_{1-x}$Mo$_x$O$_2$, a 2D-M2 phase, governed by largely 2D ordering of atomic displacements, was discovered using the new 3D-$\Delta$PDF method. We have shown that the structure is driven by geometric frustration that naturally emerges from the rutile structure. The fact that only half the vanadium ions dimerize may indicate the importance of Mott physics, but a full electronic structure treatment of the 2D-M2 phase remains to be determined. The extremely sharp diffraction features indicate that most of the correlations derive from static order, but there may be a dynamic component, which could be revealed using inelastic x-ray/neutron scattering. The 2D-M2 phase weakens with increasing Mo content, but exists at least up to $x \approx 0.3$. A systematic study of the structural parameters in detail over a wide compositional range will reveal how the fundamental order parameters evolve, which may help explain what conditions are required for the 2D-M2 state to manifest. This could result in a more complete physical model for VO$_2$ and the underlying physics universal to \textit{d}-shell-active rutile phases. 

\begin{acknowledgments}
This material is based upon work supported by the U.S. Department of Energy (DOE), Office of Science, Office of Basic Energy Sciences, Neutrons Scattering Sciences and EPSCoR under award DE-SC0018174.  Work at the Materials Science Division at Argonne National Lab was supported by the U.S. DOE, Office of Science, Office of Basic Energy Sciences, Materials Sciences and  Engineering Division. Use of the Advanced Photon Source at Argonne National Laboratory was supported by the U. S. Department of Energy, Office of Science, Office of Basic Energy Sciences, under Contract No. DE-AC02-06CH11357. Electronic transport work at UCLA was supported by NSF DMREF program under the award NSF DMREF project DMREF-1629457. During the diffraction experiments, sample environment maintenance and data collection were both aided by Doug Robinson. The authors thank Patrick LeClair and Sanjoy Sarker for helpful discussions.  
\end{acknowledgments}

\bibliography{rutilelibrary}

\end{document}